\documentclass[11pt]{article}
\pdfoutput=1
\usepackage{comment}
\usepackage[margin=2.5cm]{geometry}
\usepackage{amsmath,amssymb,extarrows,mathtools,graphicx,subfigure,setspace,bbold}
\usepackage{cite}
\usepackage{slashed}
\usepackage[dvipsnames]{xcolor}
\usepackage{hyperref}
\hypersetup{colorlinks=true, linkcolor=blue, citecolor=magenta, linktoc=page}
\numberwithin{equation}{section}
\newcommand{\be}{\begin{equation}}
\newcommand{\bea}{\begin{eqnarray}}
\newcommand{\eea}{\end{eqnarray}}
\newcommand{\ba}{\begin{align}}
\newcommand{\ea}{\end{align}}
\newcommand{\ee}{\end{equation}}

\def\ie{{\it i.e.}}

\newcommand{\Th}{\hat{T}}
\newcommand{\Tt}{\tilde{T}}

\begin{document}

\begin{titlepage}
\vspace{10mm}
\begin{flushright}
%IPM/P-2018/008 \\

\end{flushright}

\vspace*{20mm}
\begin{center}

{\Large {\bf Complexity of Hyperscaling Violating Theories  at Finite Cutoff}\\
}

\vspace*{15mm}
\vspace*{1mm}
{Mohsen Alishahiha${}^\&$  and  Amin Faraji Astaneh$^{\dagger,*}$}

 \vspace*{1cm}

{\it ${}^\&$ School of Physics,
Institute for Research in Fundamental Sciences (IPM)\\
P.O. Box 19395-5531, Tehran, Iran\\
 ${}^\dagger$ Physics Department, Faculty of Sciences, Arak University, Arak 38156-8-8349, Iran\\
 ${}^*$  School of Particles and Accelerators,\\
Institute for Research in Fundamental Sciences (IPM)\\
P.O. Box 19395-5531, Tehran, Iran
}

 \vspace*{0.5cm}
{E-mails: {\tt alishah@ipm.ir, faraji@ipm.ir}}%

\vspace*{1cm}
%%\maketitle
\end{center}

\begin{abstract}
Using the complexity equals action proposal we study holographic complexity for
 hyperscaling violating theories in the presence of a finite cutoff that, in turns, requires
to  obtain all  counter terms needed to have finite boundary energy momentum tensor. 
These terms could give non-trivial contributions to the complexity.  
We observe that having a finite UV cutoff would enforce us to have
 a cutoff behind the horizon of which the value is fixed by the UV cutoff;
 moreover, certain counter term should be defined on the cutoff behind the horizon too.
\end{abstract}

\end{titlepage}

\newpage
\tableofcontents
\noindent
\hrulefill
\onehalfspacing

\section{Introduction}\label{sec 1}

%An interesting development in the literature of theoretical high energy physics is to study 
%a conformal theory deformed by an  operator that is quadratic 
%in the stress energy tensor, known as $T\bar{T}$ deformation \cite{Zamolodchikov:2004ce}. 
 % Although typically deforming 
%a conformal field theory by an irrelevant operator would remove UV fixed point and 
%makes it non-local at high energies, it was shown that for the this special kind of deformation 
%the resultant theory is still exactly solvable\cite{{Smirnov:2016lqw},{Cavaglia:2016oda}}.
%In particular the obtained theory is UV complete and the spectrum of the deformed theory 
%can be  determined  non-perturbatively and rather in a compact form.

In the context of AdS/CFT correspondence \cite{Maldacena:1997re}  it was proposed 
that $T\bar{T}$ deformation  of a two dimensional conformal field theory
\cite{Zamolodchikov:2004ce}
has an interesting  holographic dual in terms of  an AdS$_3$ geometry  with a finite radial cutoff\cite{McGough:2016lol}, see also \cite{Caputa:2019pam}. Generalization of 
 $T\bar{T}$ deformation to higher dimensional conformal field theories has also been 
 studied in  \cite{{Taylor:2018xcy},{Hartman:2018tkw}} where it was proposed that 
 the corresponding deformation has a gravitational dual given by a higher dimensional 
 AdS (black brane) geometry with a finite radial cutoff.
 
Using gravitational description of  $T\bar{T}$ deformation of a conformal field theory,
the holographic complexity for the 
 black brane solutions of the Einstein gravity  at a finite radial cutoff was studied in
\cite{Akhavan:2018wla} where it was shown that the finite UV cutoff induces a cutoff 
behind the horizon (see also \cite{{Alishahiha:2019cib},{Hashemi:2019xeq}}). In particular 
it was shown that the presence of the behind the horizon cutoff is crucial to find the  expected 
result for the holographic complexity  of  
the Jackiw-Teitelboim gravity \cite{Alishahiha:2018swh}.\footnote{
It should be  mentioned that there is another approach  to get the desired results for 
the holographic complexity for the Jackiw-Teitelboim gravity \cite{{Brown:2018bms},
{Goto:2018iay}}. In this approach the authors have considered  the contribution of
 certain boundary term which is required to impose the Neumann boundary condition for the gauge field.
  This in fact could be naturally understood  if one considers the two dimensional  model 
as a dimensionally reduced four dimensional RN black hole. Of course since the aim 
of the present paper is to explore the role of the behind the horizon cutoff , in what follows we will
follow the approach proposed in \cite{Akhavan:2018wla}.}
 It is 
also important to note that in order to get 
the desired result one needs to consider  contributions of certain counter terms
appearing in the context of holographic renormalization that are usually  required to get finite 
on shell action.

As motivated above, in order to further illustrate the roles of the counter terms and the 
behind the horizon cutoff, in 
this paper we will  study holographic complexity for theories with hyperscaling violation at 
a finite cutoff.
We note that complexity for hyperscaling violating theories has been already studied 
in  \cite{{Swingle:2017zcd},{Alishahiha:2018tep}}. Of course in what follows the aim is
to explore the effect of a finite UV cutoff in the computations of complexity. Indeed,  we will 
see that in this case in order to get a consistent result  for the complexity  one is forced 
to have  a cutoff behind the horizon of which the value is fixed by the finite UV cutoff. 
We note also that besides this cutoff there are
 certain counter terms whose contributions  should be  taken into account too. 

To proceed, first of all,  one needs  to study the hyperscaling violating geometry
in the presence of a finite radial cutoff which might be thought of as a $T\bar{T}$-like 
generalization of non-relativistic theories.  Actually to read the energy of the system 
with a UV cutoff one needs  to fully study all possible counter terms required to get finite 
boundary energy momentum tensor. Indeed,  this is what we will do  in the present 
paper by which we will be able to compute the finite cutoff corrections to the energy.

Geometries with hyperscaling violating factor have been studied in 
\cite{{Gouteraux:2011ce},{Huijse:2011ef}}. To fix our notation, as a minimal model, we 
will consider  an Einstein-Dilaton-Maxwell theory in $d+2$ dimensions  whose action may 
be given by \cite{Alishahiha:2012qu}
\bea
\label{action}
S^{\rm bulk}\!\!&\!\!=\!\!&\!\!\frac{1}{16\pi G}\int d^{d+2}x\sqrt{-g}\left[R-\frac{1}{2}(\partial\phi)^2+V(\phi)
-\frac{1}{4} e^{\eta \phi}F^2\right],\cr &&\cr 
 S^{\rm GH}\!\!&\!\!=\!\!&\!\!\frac{1}{8\pi G}\int d^{d+1}x\sqrt{-h}K.
 \eea
 The Gibbons-Hawking action is required to have a well defined variational principle. Nonetheless 
 its presence is curtail for other purposes such as to find finite free energy. It may also have 
 non-trivial contribution to the holographic complexity. As for a potential we will  consider the 
following term
\be
V=V_0e^{\zeta\phi}.
\ee
Here $\eta, \zeta$ and $V_0$ are free parameters of the model. This model,  indeed, admits 
solutions with  hyperscaling violation and non-trivial anisotropy. In fact, the vector field is 
required to produce an anisotropy while non-trivial potential, as given above, is needed 
to have hyperscaling violating factor.  The corresponding black brane solution is 
\bea\label{solution}
ds^2\!\!&\!\!=\!\!&\!\!
r^{-2{\theta}_e}\left(-r^{2z}f(r)dt^2+\frac{dr^2}{r^2f(r)}+r^2d\vec{x}^2\right),
\;\;\;\;\;\;\;\;\;\;\;f(r)=1-\frac{r_h^{z+d_e}}{r^{z+d_e}}\,,
\cr &&\cr
 A_{t}(r)\!\!&\!\!=\!\!&\!\!
 \sqrt{\frac{2(z-1)}{z+d_e}}\;r^{z+d_e} f(r),\;\;\;\;\;\;\;\;\;\;\;\;\;\;\;\;\;\;\;\;\;\;\;\;\;\;\;\;\;\;
 \;\;\;\;\;\;\;\;\;\;\;
e^{\eta\phi}=r^{-2(d_e+{\theta}_e)}\,,
\eea
where $r_h$ is the radius of horizon. Note that in our notations the boundary is located at 
$r_\infty\rightarrow \infty$.  Here $\theta_e=\frac{\theta}{d}$ and $d_e=d-\theta$ may 
be thought of as  effective hyperscaling and dimension, respectively.  $z$ is also the 
anisotropy exponent. The parameters of the solution and those of the action are related by
\be
\eta= \frac{-2(d_e+{\theta}_e)}{\sqrt{2d_e(z-1-{\theta}_e)}},
\;\;\;\;\;\;\;\;
V_0=(z+d_e-1 ) (z+d_e  ),\;\;\;\;\;\;\;\;\zeta=\frac{{2\theta}_e}{\sqrt{2d_e(z-1-{\theta}_e)}}\,.
\ee 
Note  that there is no charge associated with the gauge field, even though there is a 
non-zero gauge field. Indeed,  its effect is just to reproduce an anisotropy and thus 
setting $z=1$ the gauge field vanishes.  This geometry provides a holographic description  
of a field theory   with hyperscaling violating symmetry in $d+1$ dimensions. 

The rest of the paper is organized as follows. In the following section we will study holographic 
energy at the finite cutoff. To do so, one will have to find certain counter terms required 
to get finite  on shell action when evaluated over the whole space time. These counter terms are
also needed to have finite energy momentum tensor. In section four using 
``complexity equals action'' proposal  we will compute complexity of hyperscaling 
violating theories with a finite cutoff. We shall see that to find a consistent result a cutoff 
behind the horizon is required whose value  is given by the finite cutoff. 
It is also important to have certain boundary terms that contribute to the complexity.
 The last section is devoted to discussions where we will also give a comment on a possible
 holographic dual description of  a $T\bar{T}$-like  deformation of a theory with hyperscaling violation.
%%%%%%%%%%%%%%%%%%%%%%%%%%%%%%%%%%%%%%%%%%
%%%%%%%%%%%%%%%%%%%%%%%%%%%%%%%%%%%%%%%%%%
%%%%%%%%%%%%%%%%%%%%%%%%%%%%%%%%%%%%%%%%%%
%%%%%%%%%%%%%%%%%%%%%%%%%%%%%%%%%%%%%%%
%%%%%%%%%%%%%%%%%%%%%%%%%%%%%%%%%%%%%%%
%%%%%%%%%%%%%%%%%%%%%%%%%%%%%%%%%%%%%%

\section{Holographic  energy  at a finite cutoff}

In this section we will compute the energy of the model when there is a finite radial cutoff.
To do so, one needs regularized boundary energy momentum tensor that, in turn,  
requires to have full action including all counter terms to make the gravitational free energy 
finite. 

Therefore in what follows we will first compute  free energy of the solution \eqref{solution}. 
To proceed we note that for the solution \eqref{solution} one has 
\be
\sqrt{-g}\left(R-\frac{1}{2}(\partial \phi)^2+V_0 
e^{\xi \phi}-\frac{1}{4}e^{\eta \phi}F^2\right)=-2(1-{\theta}_e)(d_e+z) {r^{d_e+z-1}},
\ee
leading to the following expression for the bulk part of the action
\be
S^{\rm bulk}=-\frac{ V_{d+1} (1-\theta_e)(z+d_e)}{8\pi G_N d}\int_{r_h}^{r_\infty} dr\, r^{z+d_e-1}
=(1-{\theta}_e)\frac{V_{d+1}}{8\pi G_N}\left(-r_\infty^{z+d_e}+
r_h^{z+d_e}\right),
\ee
where $r_\infty$ is a UV cutoff that will eventually be sent to infinity. On the other hand for the 
Gibbons-Hawking term one finds
\be
S^{\rm GH}=(z+d_e-{\theta}_e)\frac{ V_{d+1}}{8\pi G_N}r_\infty^{z+d_e}
-(z+d_e-{2\theta}_e)\frac{ V_{d+1}}{16\pi G_N}r_h^{z+d_e}\, .
\ee
Putting both contributions together one arrives at 
\be
S^{\rm bulk}+S^{\rm GH}=(z+d_e-1)\frac{ V_{d+1}}{8\pi G_N}r_\infty^{z+d_e}-
(z+d_e-2)\frac{ V_{d+1}}{16\pi G_N}r_h^{z+d_e}\, ,
\ee
that is divergent as $r_\infty$ approaches infinity. It is, of course, known that to remove 
all divergent terms one needs to add certain counter terms that play an important role
 in the context of holographic renormalization \cite{Skenderis:2002wp}. For the cases
 in which the model has non-zero anisotropy exponent the corresponding counter terms
 have been studied in \cite{Kiritsis:2015doa}. Motivated by this paper and to
 accommodate non-zero hyperscaling violating factor we will consider the following ansatz
 for the counter term
 \be\label{CT0}
S^{\rm ct}=-\frac{1}{16\pi G_N}\int d^{d+1}x\sqrt{-h}\,e^{\frac{1}{2}\zeta\phi}
(c_1+c_2\, e^{\eta\phi}A_\mu A^\mu)\,
\ee
where $c_1$ and $c_2$ are two numerical constants that may be fixed by requiring the finiteness 
of the free energy that produces desired entropy of the black brane. For 
the  black hole solution \eqref{solution}  the above  counter term reads
\be
S^{\rm ct}=-\frac{ V_{d+1}}{16\pi G_N}\left(c_1-\frac{ 2 c_2 (z-1)}{d_e+z}\right)r_\infty^{d_e+z}+
\frac{ V_{d+1}}{32\pi G_N}\left(c_1-\frac{6 c_2 (z-1)}{d_e+z}\right)r_h^{z+d_e}.
\ee
It is then easy to see that requiring to have finite free energy that gives correct entropy one 
should have
\be
c_1=2d_e+z-1,\;\;\;\;\;\;\;\;\;\;\;c_2=-\frac{z+d_e}{2}\,,
\ee
so that
\be
S^{\rm tot}=S^{\rm bulk}+S^{\rm GH}+S^{\rm ct}=\frac{z V_{d+1}}{16\pi G_N}r_h^{z+d_e}\, .
\ee

The next step is to check whether the above counter term is enough to have finite
(regular) boundary energy momentum tensor. To see this, one needs to compute the 
corresponding boundary energy momentum tensor for our model. Following 
\cite{Kiritsis:2015doa} one should note that there are several parts that contribute to the 
boundary energy momentum tensor which  may be decomposed as follows
\be
T^\mu_\nu=T^{({\rm b})\mu}_{\nu}+T^{({\rm c})\mu}_{\nu}+\tau^{({\rm b})\mu}_\nu
+\tau^{({\rm c})\mu}_\nu\, ,
\ee 
where $(b)$ and $(c)$ stand for ``bulk'' and ``counter term'' that indicate
whether the corresponding term  comes from bulk action (including Gibbons-Hawking term) 
or  the counter term. More  explicitly one has 
\bea
\!\!&&\!\!T^{({\rm b})\mu}_{\nu}=\frac{1}{8\pi G_N}(K^\mu_{\nu}-\delta^\mu_{\nu}K)\ ,\,\;\;\;T^{({\rm c})\mu}_{\nu}=\frac{2d_e+z-1}{16\pi G_N}e^{\frac{1}{2}\zeta\phi}\,\delta^\mu_{\nu}
+\frac{z+d_e}{16\pi G_N}e^{\eta\phi+\frac{1}{2}\zeta\phi}
(A^\mu A_\nu-\frac{1}{2}\delta^\mu_{\nu}A_\alpha 
A^\alpha)\, \nonumber,\\
\!\!&&\!\!\tau^{({\rm b})\mu}_\nu=\frac{1}{16\pi G_N}e^{\eta\phi}n_\alpha 
F^{\alpha\mu}A_\nu\, ,\;\;\;\;\;\;\;
\tau^{({\rm c})\mu}_\nu=-\frac{z+d_e}{16\pi G_N}e^{\eta\phi+\frac{1}{2}\zeta\phi}
A^\mu A_\nu\, .
\eea
In fact, the the first term is the standard Brown-York energy momentum tensor. It is worth 
noting that for the model we are considering the energy momentum tensor is not 
symmetric \cite{Kiritsis:2015doa}.  By making use of the explicit form of the black brane solution \eqref{solution} different components of the energy momentum tensor read
\bea
\!\!&&\!\!T^{(\rm b)t}_t=-\frac{d_e}{8\pi G_N}f^{\frac{1}{2}}r^{{\theta}_e}\ ,\,\;\;\;\;\;\;\;\;\;\;\;\;\;\;\;\;\;\;
\;\;\;\;\;\;\;\;\; 
T^{(\rm b)i}_j=\frac{-1}{16\pi G_N}f^{-\frac{1}{2}}[z+d_e+(z+d_e-2)f]\, 
r^{{\theta}_e}\delta^i_j\, \nonumber,\\
\!\!&&\!\!T^{(\rm c)t}_t=\frac{1}{16\pi G_N}[2d_e+z-1-(z-1)f]\,r^{{\theta}_e}\,,\;\;\;
T^{(\rm c)i}_j=\frac{1}{16\pi G_N}[2d_e+z-1+(z-1)f]r^{{\theta}_e}\delta^i_j\, \nonumber,\\
\!\!&&\!\!\tau^{(\rm b) t}_t=-\frac{z-1}{8\pi G_N}f^{1/2}r^{{\theta}_e}\,,\;\;\;\;\;\;\;\;\;\;\;\;\;\;\;\;\;\;\;\;\;\;\;
\tau^{(\rm c) t}_t=\frac{z-1}{8\pi G_N}f\,r^{{\theta}_e}\, .
\eea
Utilizing the holographic renormalization procedure \cite{Skenderis:2002wp} one may define 
the expectation value of different components of the dual field theory energy moment 
as follows
\be
\Th^\mu_\nu =\lim_{r_\infty\rightarrow\infty}r_\infty^{z+d_e-{\theta}_e}\ T^\mu_\nu\, ,
\ee
that more explicitly are given by 
\bea
\!\!&&\!\!\Th^{(\rm b)t}_t=\frac{-1}{8\pi G_N}(d_e\, r_\infty^{z+d_e}-\frac{d_e}{2}  
r_h^{z+d_e})\,,\;\;\;\Th^{(\rm c)t}_t=\frac{1}{8\pi G_N}(d_e\, r_\infty^{z+d_e}
+\frac{z-1}{2}r_h^{z+d_e})\, ,\\
\!\!&&\!\!\hat{\tau}^{({\rm b})t}_t
=\frac{-1}{8\pi G_N}((z-1)r_\infty^{z+d_e}-\frac{z-1}{2}r_h^{z+d_e})\, ,\;\;\;
\hat{ \tau}^{({\rm c})t}_t
=\frac{1}{8\pi G_N}((z-1)r_\infty^{z+d_e}-(z-1)r_h^{z+d_e})\,\nonumber ,\\
\!\!&&\!\!\Th^{({\rm b})i}_j=\frac{-1}{8\pi G_N}((z+d_e-1)r_\infty^{z+d_e}
+\frac{1}{2}r_h^{z+d_e})\delta^i_j\, ,\;\;\;\Th^{({\rm c})i}_j=\frac{1}{8\pi G_N}((z+d_e-1)r_\infty^{z+d_e}-\frac{z-1}{2}
r_h^{z+d_e})\delta^i_j\, .\nonumber
\eea
It is then easy to sum up the corresponding terms resulting in
\be
\Th^t_t =\frac{d_e}{16\pi G_N}r_h^{z+d_e} \ \ , \ \ \Th^i_j =\frac{-z }{16\pi G_N}
\delta^i_j r_h^{z+d_e}\, ,
\ee
which can be used to find the trace condition for the theory as follows
\be
z\,\Th^t_t +\frac{d_e}{d}\,  \Th^i_i  =0\, .
\ee
Therefore, the counter term we have considered is also enough to have finite energy momentum 
tensor. The same argument can be deduced for the expectation value of the operator dual to the dilaton field. It is then straightforward to compute the energy of the solution with or without cutoff. 
In particular setting the theory at a radial cutoff $r=r_c$ the corresponding energy is found
\bea
{E}\!\!&\!\!=\!\!&\!\! r_c^{z-{\theta}_e}\int d^dx\sqrt{g}\ T^t_t=\frac{V_d}{16\pi G_N}r_c^{z+d_e}
\left(\sqrt{f(r_c)}-1\right)\left(-2d_e+(z-1)(\sqrt{f(r_c)}-1\right)
\cr &&\cr
\!\!&\!\!=\!\!&\!\!\frac{(z+d_e-1) V_d}{8\pi G_N}r_c^{z+d_e}
\left(1-\sqrt{1-\frac{r_h^{z+d_e}}{r_c^{z+d_e}}}\right)-\frac{(z-1)V_d}{16\pi G_N}
r_h^{z+d_e}, 
\eea
which  may be 
recast into the following form
\bea\label{ECT}
{E}=\frac{d_eV_d}{16\pi G_N}
r_h^{z+d_e}+ \frac{(d_e+z-1)V_d }{16\pi G_N}r_c^{z+d_e}
\left(1-\sqrt{1-\frac{r_h^{z+d_e}}{r_c^{z+d_e}}}\right)^2\,.
\eea 
This is indeed the final form of the energy at a finite cutoff that will be used when we 
want to study  the late time behavior of the complexity. The same behavior of the energy can be seen in \cite{Cardy} where the $T\bar{T}$ deformation of a two dimensional Lifshitz theory has been studied.   

It is worth noting that in the large cutoff limit, $r_c\rightarrow \infty$, the second term in the 
above expression vanishes leaving us with the first term that is, indeed, the mass of the black 
hole solution \eqref{solution}.

%%%%%%%%%%%%%%%%%%%%%%%%%%%%%%%%%%%%%%%%%%%%%%
%%%%%%%%%%%%%%%%%%%%%%%%%%%%%%%%%%%%%%%%%%%%%%
%%%%%%%%%%%%%%%%%%%%%%%%%%%%%%%%%%%%%%%%%%%%%%
%%%%%%%%%%%%%%%%%%%%%%%%%%%%%%%%%%%%%%%%%%%%%%
%%%%%%%%%%%%%%%%%%%%%%%%%%%%%%%%%%%%%%%%%%%%%%
%%%%%%%%%%%%%%%%%%%%%%%%%%%%%%%%%%%%%%%%%%%%%%

%%%%%%%%%%%%%%%%%%%%%%%%%%%%%%%%%%%%%%%%%%%%%%
%%%%%%%%%%%%%%%%%%%%%%%%%%%%%%%%%%%%%%%%%%%%%%
%%%%%%%%%%%%%%%%%%%%%%%%%%%%%%%%%%%%%%%%%%%%%%
%%%%%%%%%%%%%%%%%%%%%%%%%%%%%%%%%%%%%%%%%%%%%%
%%%%%%%%%%%%%%%%%%%%%%%%%%%%%%%%%%%%%%%%%%%%%%
%%%%%%%%%%%%%%%%%%%%%%%%%%%%%%%%%%%%%%%%%%%%%%

\section{Complexity at finite cutoff}

\begin{figure}
\begin{center}
\includegraphics[scale=1]{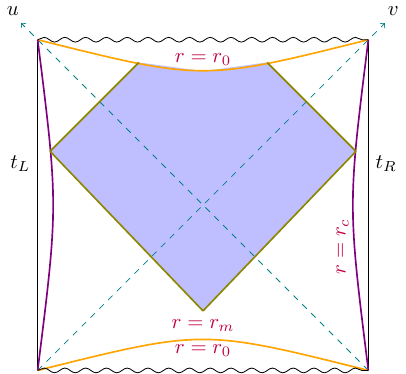}
\end{center}
\caption{The  WDW patch for the theory at finite cutoff $r_c$. There is also a cutoff behind the 
horizon, $r_0$ of which the value is fixed by UV cutoff $r_c$. \label{fig1}}
\end{figure}

In this section we would like to compute the holographic complexity for the hyperscaling 
violating geometry with a  finite cutoff denoted by $r_c$ as shown in the figure \ref{fig1}. 
The holographic complexity for such geometries have been studied in \cite{{Swingle:2017zcd},
{Alishahiha:2018tep}} where it was shown that at the late time the complexity growth approaches
the following constant 
\be\label{LTB}
\frac{d {\cal C}}{d\tau}=\frac{d_e+z-1}{d_e} \, 2 E_0,
\ee 
where $E_0$, the energy of the back brane, is given by 
\be
E_0=\frac{d_eV_d}{16\pi G_N} r_h^{d+z}.
\ee
Although it violates the naive Lloyd's bound given by the twice of the energy
\cite{Lloyd:2000}, it is still given by 
a constant related to the energy of the system. Having put  the system at a finite cutoff one 
would expect that the late time behavior of complexity should be given in terms of the 
deformed energy \eqref{ECT} as follows
\be\label{LTBc}
\frac{d {\cal C}}{d\tau}=\frac{d_e+z-1}{d_e} \, 2 E.
\ee 
This is what we would like to explore in this section using ``complexity=action'' proposal  (CA)
for the holographic complexity.  According to this proposal the quantum computational 
complexity of a   holographic state is given by the on-shell action evaluated on a bulk 
region known as the  ``Wheeler-De Witt'' (WDW) patch \cite{Brown:2015bva, Brown:2015lvg}
 \be
{\cal C}(\Sigma)=\frac{I_{\rm WDW}}{\pi \hbar}.
\ee
Here the WDW patch is defined as the domain of dependence of any Cauchy surface in the bulk 
whose intersection with the asymptotic boundary is the time slice $\Sigma$. 
 
Based on the CA proposal, we evaluate 
the on shell action of the WDW patch shown in the figure \ref{fig1} where the WDW patch is 
restricted by the UV cutoff $r_c$. For further use we have also set another cutoff near 
singularity at $r=r_0$. In principle, one might naively expect that the cutoff $r_0$ can be 
sent to zero at the end of the day. As we will see this is not the case and indeed 
our ultimate goal is to explore the role of this cutoff.

It is worth mentioning that  the on shell action evaluated on a WDW patch gets 
different contributions from different parts of the action including bulk, boundary and 
corner parts \cite{{Parattu:2015gga},{Parattu:2016trq},{Lehner:2016vdi},{Carmi:2017jqz}}. 
In what follows  we will compute each conurbation when the corresponding WDW 
patch is bounded by two cutoffs: $r_c$ and $r_0$.

To proceed let us fix our notations for the boundary and the joint points at first.  Two null boundaries of the 
corresponding WDW patch terminating to the joint point $r_m$ are given by
\be
t=t_R-r^*(r_c)+r^*(r),\;\;\;\;\;\;\;\;\;t=-t_L+r^*(r_c)-r^*(r),
\ee
by which the joint point $r_m$ is given by $\tau=2(r^*(r_c)-r^*(r))$, with
$r^*(r)=\int \frac{dr}{r^{z+1} f(r)}$. Here $t_L (t_R)$, is  time coordinate of left (right)
boundary located at the cutoff surface  $r=r_c$ and, the boundary time is defined by 
$\tau=t_L+t_R$. The  null vectors associated with these null boundaries are also given by
\be\label{NV}
k_1=\alpha \left(\partial_t+\frac{\partial_r}{r^{z+1} f(r)}\right), \;\;\;\;\;\;\;\;\;\;\;\;\;\;\;
k_2=\beta \left(-\partial_t+\frac{\partial_r}{r^{z+1} f(r)}\right),
\ee
where $\alpha$ and $\beta$ are two free parameters appearing due to ambiguity associated 
to the definition of norms of the null vectors. We note also that in our notation we have a space like 
boundary at $r=r_0$ of which the normal vector is given by 
$n=\frac{\partial_r}{r_0^{\theta_e+1}\sqrt{|f(r_0)|}}$.

To compute the on shell action let us start from the bulk part that is essentially given by bulk action
given in equation \eqref{action} evaluated for the solution \eqref{solution}. The bulk
contribution to the on shell action is
\bea
S^{\rm bulk}\!&\!=\!&\!-\frac{V_d}{2\pi G_N}(1-{\theta}_e)(d_e+z)\int_{r_{0}}^{r_c} 
{dr}\,{r^{d_e+z-1}}\left({r^*(r_c)-r^*(r)}\right)\cr &&\cr
&&-\frac{V_d}{4\pi G_N}(1-{\theta}_e)(d_e+z)\int_{r_{0}}^{r_{m}} 
{dr}\,{r^{d_e+z-1}}\left({\frac{\tau}{2}-r^*(r_c)+r^*(r)}\right),
\eea
that might be simplified to find
\be
S^{\rm bulk}=\frac{(1-\theta_e)V_d}{4\pi G}
\left(\frac{r_m^{d_e}+r_0^{d_e}-2r_c^{d_e}}{d_e}
+\left(\frac{\tau}{2}+r^*(r_c)-r^*(r_0)\right)r_0^{d_e+z}f(r_0)\right)\,.
\ee

In the next step we consider the contribution of the joint points. In the present case 
there are five corners two of which at UV cutoff, the other two at the  behind the horizon cutoff 
and the last one at the joint point $r_m$. The action of a joint point has the following general  form
\cite{Lehner:2016vdi}
\be
\pm \frac{1}{8\pi G}\int d^dx\, d\lambda \sqrt{\gamma} \,\log a
\ee
where $a$ is the inner product of the  normal vectors associated with two corresponding 
intersecting  boundaries (there is also a factor of one-half if both boundaries are null). Here
$\gamma$ is determinant of the induced metric on the joint points and $\lambda$ 
 is the null coordinate defined on the null segments that we choose to be Affine. The ``$\pm$''
 sings tell us which corner contribution is being computed (for more details see
 \cite{Lehner:2016vdi}).  In fact using the above normal vectors given in the equation  
 \eqref{NV} one may compute the contribution of corner points 
 \bea
S^{\rm joint}\!\!&\!\!=\!\!&\!\!
\frac{V_d}{8\pi G} (z-\theta_e)\left(2 r_c^{d_e}\log r_c^2-r_m^{d_e}\log r_m^2
-r_0^{d_e}\log r_0^2\right)+\frac{V_d}{8\pi G}(r_m^{d_e}+r_0^{d_e}-2r_c^{d_e})\log \alpha\beta
\cr &&\cr
\!\!&\!\!+\!\!&\!\!
\frac{V_d}{8\pi G}\left(2r_c^{d_e}\log |f(r_c)|-r_m^{d_e}\log |f(r_m)|-r_0^{d_e}\log |f(r_0)|\right)\,.
\eea
From the second term it is clear that the on shell action suffers 
from an ambiguity associated with the definition of null vectors as mentioned above. 
Indeed, in order  to remove this  ambiguity one needs to add extra counter terms 
associated to each null  boundaries. The corresponding counter term is 
\cite{{Lehner:2016vdi},{Reynolds:2016rvl}}
\be
\pm \frac{1}{8\pi G}\int d\lambda d^dx\sqrt{\gamma}\,\Theta\log\frac{|\Theta|}{d_e}
\ee
which should be evaluated on all null boundaries with a proper sign (see \cite{Lehner:2016vdi} for more 
details). Here $\Theta$ is defined by 
\be
\Theta=\frac{1}{\sqrt{\gamma}}\frac{\partial\sqrt{\gamma}}{\partial\lambda},
\ee
 It is 
 then straightforward to compute the contribution of this term for all null boundaries
\bea
S^{\rm ct}_1\!\!&\!\!=\!\!&\!\!
\frac{V_d}{4\pi G}\frac{z-2\theta_e}{d_e}(2r_c^{d_e}-r_m^{d_e}-r_0^{d_e})+
\frac{V_d}{8\pi G}(2r_c^{d_e}-r_m^{d_e}-r_0^{d_e})\log\alpha\beta\cr &&\cr
&+&\frac{V_d}{8\pi G}(2\theta_e-z)\left(2r_c^{d_e}\log r_c^2-r_m^{d_e}\log r_m^2
-r_0^{d_e}\log r_0^2\right)\,.
\eea
It is then evident that the ambiguous term drops from the on shell action.

The WDW patch we are considering has a space like boundary at $r=r_0$ and therefore 
one should also consider a Gibbons-Hawking term on this boundary of which the contribution is
\be
S^{\rm GH}=-\frac{V_d}{4\pi G}\left({(d_e+z-{\theta}_e)}r_0^{d_e+z}-\frac{(d_e+z-2
{\theta}_e)}{2}
r_h^{d_e+z}\right)\left(\frac{\tau}{2}+r^*(r_c)-r^*(r_0)\right)\,.
\ee

At this point it is important to emphasis that by on shell action we mean to consider 
contributions of all terms needed to have  a general covariant action with a well
imposed variational principle that result in a finite on shell action. Actually, the terms 
we have considered so far are enough to maintain the first two conditions, thought the 
resultant  on shell action is still divergent. Therefore one should add  other boundary terms 
whose contributions remove the divergences of the on shell action leading to a finite value,
 while keeping  the variational principle unaffected. Of course these terms may 
also contribute  to the finite value of the on shell action. Indeed, for the model under 
consideration such a counter  term has been given in \cite{Alishahiha:2018tep} that is
\be
S^{\rm ct}_2=\pm \frac{1}{8\pi G}\int d\lambda d^dx \sqrt{\gamma}\,\Theta\left(\frac{1}{2}\zeta \phi +\frac{z-1}{d_e}\right),
\ee
whose contribution to the on shell action when evaluated for all null boundaries is
\be
S^{\rm ct}_2=-\frac{V_d}{4\pi G}\frac{z-\theta-1}{d_e}(2r_c^{d_e}-r_m^{d_e}-r_0^{d_e})
-\frac{V_d\theta_e}{8\pi G}\left(2r_c^{d_e}\log r_c^2-r_m^{d_e}\log r_m^2-r_0^{d_e}\log r_0^2
\right).
\ee
Now putting all terms we have computed so far together one arrives at
\bea
S^{\rm T}\!\!&\!\!=\!\!&\!\!S^{\rm bulk}+S^{\rm joint}+S^{\rm GH}+S^{\rm ct}_1+S^{\rm ct}_2
=-
\frac{V_d}{8\pi G}\left(r_m^{d_e}\log |f(r_m)|+r_0^{d_e}\log |f(r_0)|\right)\\
\!\!&\!\!\!\!&\!\!-\frac{(d_e+z-1)V_d}{4\pi G} r_0^{d_e+z}
\left(\frac{\tau}{2}+r^*(r_c)-r^*(r_0)\right)
+\frac{(d_e+z-2)V_d}{8\pi G}r_h^{d_e+z}
\left(\frac{\tau}{2}+r^*(r_c)-r^*(r_0)\right).\nonumber
\eea
It is then easy to compute the action growth (complexity growth) 
 \be\label{eq 3.18}
\frac{d S^{\rm T}}{d\tau}=\frac{(d_e+z-1)V_d}{8\pi G} r_h^{d_e+z}\left(1
+\frac{d_e}{2(d_e+z-1)}\tilde{f}(m)\log|f(r_m)|\right)-\frac{(d_e+z-1)V_d}{8\pi G} r_0^{d_e+z},
\ee
where
\be
\tilde{f}(r_m)=\frac{r_m^{d_e+z}}{r_h^{d_e+z}}-1\,.
\ee
Note that to reach this expression we have used the fact that  
$\frac{dr_m}{d\tau}=-\frac{1}{2}r_m^{z+1}f(r_m)$. Therefore at the late time one gets
\be
\frac{d S^{\rm T}}{d\tau}=\frac{(d_e+z-1)V_d}{8\pi G} r_h^{d_e+z}-\frac{(d_e+z-1)V_d}{8\pi G} r_0^{d_e+z},
\ee
that reduces to \eqref{LTB} for $r_0\rightarrow 0$.  Note that the time $\tau$ is defined 
as the proper time at the cutoff surface $r_c$ where the corrected energy is evaluated.
\footnote{It is worth mentioning  that since the second term between parentheses in 
the equation \eqref{eq 3.18} is  positive, the rate of change of complexity approaches the late time limit from above, violating the Lloyd's bound.}

This observation that the late time behavior is independent of the finite UV cutoff 
looks, indeed, counter intuitive.
In fact, one would expect that the late time behavior of the complexity should be controlled by 
the conserved charges of the theory (such as energy) that, in general,  are sensitive 
to the finite UV cutoff as we demonstrated in the previous section.

Moreover, as it is clear from equation \eqref{ECT} setting a finite cutoff increases 
 the energy while the complexity evaluated by the one shell action either remains 
 unchanged (for $r_0\rightarrow 0$) or  decreases for finite $r_0$, that is puzzling as well.

Therefore the conclusion could  be that either the on shell action evaluated in the 
WDW patch does not compute the complexity, or ir does but its late times behavior 
is not determined by the Lloyd's bound (physical charges at the boundary). Another 
possibility could be that there are other terms whose contributions  are missed in the computations we have done so far. 
 In what follows, in accordance with \cite{{Akhavan:2018wla},{Alishahiha:2019cib},{Hashemi:2019xeq}} 
 we assume that the CA proposal for complexity is correct and indeed one needs more 
 terms to consider.
 
 In fact, a remedy to resolve this puzzle is to further add another counter term on the behind the
 horizon cutoff. The corresponding term is 
\be
S^{\rm ct}_3= \frac{1}{8\pi G}\int d^dx \,dt\, \sqrt{|h|}\,e^{\frac{1}{2}\zeta\phi}\,(d_e+z-1).
\ee
This is indeed the counter term needed to remove the divergence associated with the space time
volume. Actually it has the same form as \eqref{CT0} when the contribution of vector 
field $A_\mu$ is excluded. A reason one may argue in favor of this term is as follows. 
Since we assume that  the vector field  vanishes at the horizon, what  remains behind the 
horizon is just the metric and therefore we will have to add a counter term that would take 
care of the metric. This is indeed what we have written above. This term  leads to the 
following new contribution to the on shell action
\be
S^{\rm ct}_3=\frac{(d_e+z-1)V_d}{4\pi G} r_0^{d_e+z}\sqrt{\frac{r_h^{d_e+z}}{r_0^{d_e+z}}-1}\;\;
\left(\frac{\tau}{2}+r^*(r_c)-r^*(r_0)\right).
\ee
Adding the contribution of this term to the on shell action at the late time limit (complexity) 
one gets\footnote{ With this additional term the complexity still violates the Lloyd's bound 
when evaluated by the corrected energy.} 
\be
\frac{d S^{\rm T}}{d\tau}=\frac{(d_e+z-1)V_d}{8\pi G} r_h^{d_e+z}+\frac{(d_e+z-1)V_d}{8\pi G}
 r_0^{d_e+z}\left(\sqrt{\frac{r_h^{d_e+z}}{r_0^{d_e+z}}-1}-1\right),
\ee
that should be compared with the equation \eqref{LTBc}. Doing so, 
one arrives at
\be
 r_0^{d_e+z}\left(\sqrt{\frac{r_h^{d_e+z}}{r_0^{d_e+z}}-1}-1\right)
 = \frac{d_e+z-1 }{d_e}r_c^{z+d_e}
\left(1-\sqrt{1-\frac{r_h^{z+d_e}}{r_c^{z+d_e}}}\right)^2\,,
\ee 
that for large $r_c$ limit reduces to
\be
r_0\approx \left(\frac{d_e+z-1}{4d_e}\right)^{\frac{2}{d_e+z}}\,\frac{r_h^3}{r_c^2}\,.
\ee
This has the same form as that for black brane solution in Einstein gravity \cite{Akhavan:2018wla}.
Therefore we would like to conclude that having set a  UV cutoff would automatically fix a 
cutoff behind the horizon.  
%%%%%%%%%%%%%%%%%%%%%%%%%%%%%%%%%%%%%%%%%%%%%%
%%%%%%%%%%%%%%%%%%%%%%%%%%%%%%%%%%%%%%%%%%%%%%
%%%%%%%%%%%%%%%%%%%%%%%%%%%%%%%%%%%%%%%%%%%%%%
%%%%%%%%%%%%%%%%%%%%%%%%%%%%%%%%%%%%%%%%%%%%%%
%%%%%%%%%%%%%%%%%%%%%%%%%%%%%%%%%%%%%%%%%%%%%%
%%%%%%%%%%%%%%%%%%%%%%%%%%%%%%%%%%%%%%%%%%%%%%

\section{Discussions}

In this paper we have studied holographic complexity for geometries with hyperscaling 
violating factor with a finite radial cutoff. We have observed that setting a finite UV cutoff 
would enforce us to have a cutoff behind the horizon of which the 
value is fixed by the UV cutoff. {{It is 
worth noting that for $\theta\rightarrow 0$ and $z\rightarrow 1$, (\ie \, in the relativistic limit) we 
recover all of the results previously reported in \cite{Akhavan:2018wla} and thus our current 
results,}} indeed, confirm our previous studies presented in \cite{Akhavan:2018wla} in which 
the same question  has been addressed for black brane solutions of Einstein gravity. 
Of course,  to reach the desired result it was  crucial to consider  the contribution
of all counter terms. In particular, we have seen that certain counter term must be added on the
 behind the horizon cutoff.  
 
 In course of study the complexity at the finite cutoff we had to compute boundary energy
 momentum tensor at a finite cutoff which, in turns, required to find all counter terms to
 get finite free energy that also results in the desired entropy of black brane.

 Following \cite{McGough:2016lol} one may expect to have  a  possible interpretation of setting a 
 redial cutoff in the gravity side in terms of a $T\bar{T}$ like deformation of the dual theory.
Actually as we have already mentioned for the model under consideration the trace
condition is given by  
\be
z\Th^t_t+\frac{d_e}{d}\Th^i_i=0,
\ee 
indicating that the model enjoys some certain scaling symmetry that is known as hyperscaling  
violating symmetry. Let us now put the theory at 
a finite cutoff in which one would expect that the corresponding  cutoff  breaks the scaling 
symmetry leading to a non-vanishing  trace condition. Of course, this can be seen from 
explicit expressions we have for the energy momentum tensor of the solution \eqref{solution}.
In fact, at leading order one has   
\be
z\Tt^t_t+\frac{d_e}{d}\Tt^i_i=-\frac{4\pi G}{r_c^{d_e+z}} {(d_e+z)(d_e+1-z)}
\frac{r_h^{2(d_e+z)}}{(16\pi G)^2}\propto \frac{E^2}{r_c^{d_e+z}}\,.
\ee 
 Here the energy momentum tensor of the cutoff theory, $\Tt$, is defined at the radial cutoff
  $r_c$ as follows
\be
\Tt_\mu^\nu=r_c^{d_e+z-\theta_e} T_\mu^\nu\,,
\ee
that clearly reduces to $\Th$ as the cutoff approaches infinity: $r_c\rightarrow \infty$. It is then 
interesting to investigate whether the right hand side of the trace condition can be written 
in terms of the energy momentum tensor itself. Actually the fact that the right hand 
side is proportional to energy squared indicates that such an expectation  might be reasonable.
%In fact using the explicit form of the energy momentum tensor for our solution 
%\bea
%&&\Tt^t_t=\frac{-1}{16\pi G}(\sqrt{f(r_c)}-1) \left(2d_e-(\sqrt{f(r_c)}-1) (z-1)\right) r_c^{{d_e}+z}\cr
%&&\Tt^i_j=\frac{1}{16\pi G}(\sqrt{f(r_c)}-1) \left(\frac{d_e+z}{\sqrt{f(r_c)}}+\sqrt{f(r_c)} (z-1)+
%1-d_e\right) r_c^{d_e+z}\,\delta^i_j\,,
%\eea
%it is straightforward to see that for arbitrary 
%$\theta$ and in the case where $z\approx 1$ one has
%\be
%z\Tt^t_t+\frac{d_e}{d}\Tt^i_i=-\frac{4\pi G}{r_c^{d_e+z}} \,\left[\Tt^{tt}\Tt_{tt}+\frac{d_e}{d}
% \Tt^{ij}\Tt_{ij}-\frac{1}{d_e}
%\left(\Tt^t_t+\frac{d_e}{d}\Tt^i_i\right)^2\right]+{\cal O}(z-1)\,,
%\ee 
%that is reminiscent of $T{\bar T}$ deformation of CFT studied in the literature. Note that in this 
%expression all terms are evaluated at the cutoff surface $r=r_c$. Therefore at least for $z=1$ 
%the cutoff geometry can be thought of as gravitational  dual of $T{\bar T}$ deformation of a 
%theory with isotropic hyperscaling violation. 

%On  the other hand when  $z\neq 1$  one would expect that $T{\bar T}$ deformation should be combined by a deformation generated by the gauge field. 
To explore this point
let us compute the vector current associated with the gauge field $A$. The corresponding 
current may be written as  $J^\mu=J^{({\rm b})\mu}
 +J^{({\rm c})\mu}$ with
 \bea
 J^{({\rm b})\mu}\!\!&\!\!=\!\!&\!\!\frac{1}{16\pi G}e^{\eta\phi} n_\alpha F^{\alpha\mu}=
 -\frac{\sqrt{2(z-1)(d_e+z)}}{16\pi G}\,\frac{1}{\sqrt{f}}\, r^{\theta_e-z-d_e}\,\delta^\nu_t\cr &&\cr
  J^{({\rm c})\mu}\!\!&\!\!=\!\!&\!\!-\frac{d_e+z}{16\pi G} e^{\eta\phi+\frac{1}{2}\zeta\phi}A^\mu=
  \frac{\sqrt{2(z-1)(d_e+z)}}{16\pi G}\, r^{\theta_e-z-d_e}\,\delta^u_t.
  \eea
Thus one gets 
\be
{\tilde J}^t=\frac{1}{16\pi G} \sqrt{\frac{2(z-1)}{d_e+z}} (\sqrt{f(r_c)}-1)\,
\frac{d_e+z}{\sqrt{f(r_c)}}\,.
\ee
Note that the current  vanishes at $r_c\rightarrow \infty$ indicating that there is no actual
charge associated with the gauge field. In fact, as we already mentioned the gauge field 
was required to produce anisotropy.  Nonetheless one could still treat the gauge field as 
a charged field and try to find the corresponding deformation using the charged black branes
case studied in \cite{{Taylor:2018xcy},{Hartman:2018tkw}}. Motivated by these works 
one could see that the following equation holds at least for the solution \eqref{solution}
\bea
z\Tt^t_t+\frac{d_e}{d}\Tt^i_i\!\!&\!\!=\!\!&\!\!-\frac{8 \pi  {d_e} G}{z (2  {d_e}+z-1)}\;\frac{1}{r_c^{d_e+z}} \,\bigg[z\Tt^{t}_{t}\Tt^{t}_{t}+\frac{d_e}{d} \Tt^{i}_{j}\Tt_{i}^{j}
-\frac{1}{d_e}
\left(z\Tt^t_t+\frac{d_e}{d}\Tt^i_i\right)^2\\ &&\cr &&
\;\;\;\;\;\;\;\;\;\;\;\;\;\;\;\;\;\;\;\;\;\;\;
-\frac{z( z-1)}{2 d_e^2} \left(z \Tt_t^t+\frac{d_e}{d}\Tt^i_i \right)\tilde{J}^t A_t+\frac{z(d_e+z)(d_e+z-1)}{2d^2_e}  \Tt^t_t\tilde{J}^t A_t\bigg].\nonumber
\eea 
Therefore we would like to propose that the 
hyperscaling violating geometries with finite radial cutoff will provide gravitational 
descriptions for non-relativistic theories deformed by  particular operators as those 
written in the right hand side of the above equation. In other words one has
\bea
\frac{\partial S}{\partial\mu}\!\!&\!\!=\!\!&\!\!\int d^{d+1}x\sqrt{h}\,\bigg[z\Tt^{t}_{t}\Tt^{t}_{t}+\frac{d_e}{d} \Tt^{i}_{j}\Tt_{i}^{j}-\frac{1}{d_e}
\left(z\Tt^t_t+\frac{d_e}{d}\Tt^i_i\right)^2
-\frac{z( z-1)}{2 d_e^2} \left(z \Tt_t^t+\frac{d_e}{d}\Tt^i_i \right)\tilde{J}^t A_t
\cr &&\cr &&\;\;\;\;\;\;\;\;\;\;\;\;\;\;\;\;\;\;\;\;\;\;
+\frac{z(d_e+z)(d_e+z-1)}{2d^2_e}  \Tt^t_t\tilde{J}^t A_t\bigg],
\eea 
where $\mu$ is the deformation parameter. Further exploration of this point would be interesting. 

Finally, we would like to comment on the alternative recipe for the holographic 
complexity, \ie\, the complexity equals volume (CV) proposal. According to this proposal one 
should extend the time constant slice of the boundary into the bulk and compute its extremized volume. Since 
the dynamical exponent comes with $tt$ component of the metric, such a quantity is blind to the 
Lifshitz exponent. Therefore in order to explore all aspects of a non-relativistic theory we chose to 
work with CA proposal instead. Moreover in the CV proposal it is known that the Einstein-Rosen
bridge approaches a maximal surface at the late time \cite{Stanford:2014jda}
that prevents us from probing the cutoff behind the horizon.  

% course it is important to emphasis once again that this equation is written on shell for the 
%particular solution we have considered in this paper.  

%In terms of this parameter the modified energy
%may be written as
%\bea
%{E}=E_0+ \frac{d_e+z-1 }{ 2d_e+z-1 }\frac{V_d d_e}
%{2z \mu}
%\left(1-\sqrt{1-\frac{2  {d_e}+z-1}{d_e}\frac{2z \mu}{V_d d_e} E_0}\right)^2\,.
%\eea 

%%%%%%%%%%%%%%%%%%%%%%%%%%%%%%%%%%%%%%%%%%%%%%
%%%%%%%%%%%%%%%%%%%%%%%%%%%%%%%%%%%%%%%%%%%%%%
%%%%%%%%%%%%%%%%%%%%%%%%%%%%%%%%%%%%%%%%%%%%%%

\subsection*{Acknowledgements}
The authors would like to kindly thank A. Akhavan, M. H. Halataei, M.R. Mohammadi,  A. Naseh, 
M.H. Vahidinia, F. Omidi and  M. R. Tanhayi for discussions on related topics. M. A. would also
like to thank ICTP for hospitality. A.F.A would also like to thank Institut des Hautes Études Scientifiques (IHES), Université Paris-Saclay, for warm hospitality.


\begin{thebibliography}{} 

%\cite{Maldacena:1997re}  
\bibitem{Maldacena:1997re} 
  J.~M.~Maldacena,
  ``The Large N limit of superconformal field theories and supergravity,''
  Int.\ J.\ Theor.\ Phys.\  {\bf 38}, 1113 (1999)
  [Adv.\ Theor.\ Math.\ Phys.\  {\bf 2}, 231 (1998)]
  [hep-th/9711200].
  %%CITATION = HEP-TH/9711200;%%
  %11018 citations counted in INSPIRE as of 11 sept. 2015


%\cite{Zamolodchikov:2004ce}\cite{McGough:2016lol}
\bibitem{Zamolodchikov:2004ce} 
  A.~B.~Zamolodchikov,
  ``Expectation value of composite field T anti-T in two-dimensional quantum field theory,''
  hep-th/0401146.
  %%CITATION = HEP-TH/0401146;%%
  %45 citations counted in INSPIRE as of 25 Oct 2018


%%\cite{Smirnov:2016lqw}
%\bibitem{Smirnov:2016lqw} 
 % F.~A.~Smirnov and A.~B.~Zamolodchikov,
  %``On space of integrable quantum field theories,''
 % Nucl.\ Phys.\ B {\bf 915}, 363 (2017)
 % doi:10.1016/j.nuclphysb.2016.12.014
 % [arXiv:1608.05499 [hep-th]].
  %%%CITATION = doi:10.1016/j.nuclphysb.2016.12.014;%%
 % %46 citations counted in INSPIRE as of 26 Oct 2018

%\cite{Cavaglia:2016oda}
%\bibitem{Cavaglia:2016oda} 
 % A.~Cavaglià, S.~Negro, I.~M.~Szécsényi and R.~Tateo,
 % ``$T \bar{T}$-deformed 2D Quantum Field Theories,''
 % JHEP {\bf 1610}, 112 (2016)
 % doi:10.1007/JHEP10(2016)112
  %[arXiv:1608.05534 [hep-th]].
  %%CITATION = doi:10.1007/JHEP10(2016)112;%%
  %70 citations counted in INSPIRE as of 17 May 2019



%\cite{McGough:2016lol}
\bibitem{McGough:2016lol} 
  L.~McGough, M.~Mezei and H.~Verlinde,
  ``Moving the CFT into the bulk with $ T\overline{T} $,''
  JHEP {\bf 1804}, 010 (2018)
  doi:10.1007/JHEP04(2018)010
  [arXiv:1611.03470 [hep-th]].
  %%CITATION = doi:10.1007/JHEP04(2018)010;%%
  %45 citations counted in INSPIRE as of 25 Oct 2018

%\cite{Caputa:2019pam}
\bibitem{Caputa:2019pam} 
  P.~Caputa, S.~Datta and V.~Shyam,
  ``Sphere partition functions \& cut-off AdS,''
  JHEP {\bf 1905}, 112 (2019)
  doi:10.1007/JHEP05(2019)112
  [arXiv:1902.10893 [hep-th]].
  %%CITATION = doi:10.1007/JHEP05(2019)112;%%
  %7 citations counted in INSPIRE as of 03 Jun 2019

%\cite{Taylor:2018xcy}{Hartman:2018tkw}
\bibitem{Taylor:2018xcy} 
M.~Taylor,
``TT deformations in general dimensions,''
arXiv:1805.10287 [hep-th].


%\cite{Hartman:2018tkw}
\bibitem{Hartman:2018tkw} 
  T.~Hartman, J.~Kruthoff, E.~Shaghoulian and A.~Tajdini,
  ``Holography at finite cutoff with a $T^2$ deformation,''
  arXiv:1807.11401 [hep-th].
  %%CITATION = ARXIV:1807.11401;%%
  %2 citations counted in INSPIRE as of 25 Oct 2018


%\cite{Akhavan:2018wla}{Alishahiha:2018swh}{Alishahiha:2019cib}{Hashemi:2019xeq}
 \bibitem{Akhavan:2018wla}
  A.~Akhavan, M.~Alishahiha, A.~Naseh and H.~Zolfi,
  ``Complexity and Behind the Horizon Cut Off,''
  JHEP {\bf 1812} (2018) 090
  doi:10.1007/JHEP12(2018)090
  [arXiv:1810.12015 [hep-th]].
  %%CITATION = doi:10.1007/JHEP12(2018)090;%%
  %8 citations counted in INSPIRE as of 19 Mar 2019
  


%\cite{Alishahiha:2019cib}{Hashemi:2019xeq}
\bibitem{Alishahiha:2019cib}
  M.~Alishahiha, K.~Babaei Velni and M.~R.~Tanhayi,
  ``Complexity and Near Extremal Charged Black Branes,''
  arXiv:1901.00689 [hep-th].
  %%CITATION = ARXIV:1901.00689;%%
  %3 citations counted in INSPIRE as of 19 Mar 2019

%\cite{Hashemi:2019xeq}
\bibitem{Hashemi:2019xeq}
  S.~S.~Hashemi, G.~Jafari, A.~Naseh and H.~Zolfi,
  ``More on Complexity in Finite Cut Off Geometry,''
  arXiv:1902.03554 [hep-th].
  %%CITATION = ARXIV:1902.03554;%%
  %2 citations counted in INSPIRE as of 19 Mar 2019

%\cite{Alishahiha:2018swh}{Alishahiha:2019cib}{Hashemi:2019xeq}
\bibitem{Alishahiha:2018swh}
  M.~Alishahiha,
  ``On Complexity of Jackiw-Teitelboim Gravity,''
  arXiv:1811.09028 [hep-th].
  %%CITATION = ARXIV:1811.09028;%%
  %5 citations counted in INSPIRE as of 19 Mar 2019  
  
 %\cite{Brown:2018bms}{Goto:2018iay}
 \bibitem{Brown:2018bms} 
  A.~R.~Brown, H.~Gharibyan, H.~W.~Lin, L.~Susskind, L.~Thorlacius and Y.~Zhao,
  ``Complexity of Jackiw-Teitelboim gravity,''
  Phys.\ Rev.\ D {\bf 99}, no. 4, 046016 (2019)
  doi:10.1103/PhysRevD.99.046016
  [arXiv:1810.08741 [hep-th]].
  %%CITATION = doi:10.1103/PhysRevD.99.046016;%%
  %11 citations counted in INSPIRE as of 24 Mar 2019
 
  

%\cite{Goto:2018iay}
\bibitem{Goto:2018iay} 
  K.~Goto, H.~Marrochio, R.~C.~Myers, L.~Queimada and B.~Yoshida,
  ``Holographic Complexity Equals Which Action?,''
  JHEP {\bf 1902}, 160 (2019)
  doi:10.1007/JHEP02(2019)160
  [arXiv:1901.00014 [hep-th]].
  %%CITATION = doi:10.1007/JHEP02(2019)160;%%
  %10 citations counted in INSPIRE as of 24 Mar 2019



%\cite{Swingle:2017zcd}{Alishahiha:2018tep}
\bibitem{Swingle:2017zcd} 
  B.~Swingle and Y.~Wang,
  ``Holographic Complexity of Einstein-Maxwell-Dilaton Gravity,''
  JHEP {\bf 1809}, 106 (2018)
  doi:10.1007/JHEP09(2018)106
  [arXiv:1712.09826 [hep-th]].
  %%CITATION = doi:10.1007/JHEP09(2018)106;%%
  %41 citations counted in INSPIRE as of 19 Mar 2019

%\cite{Alishahiha:2018tep}
\bibitem{Alishahiha:2018tep} 
  M.~Alishahiha, A.~Faraji Astaneh, M.~R.~Mohammadi Mozaffar and A.~Mollabashi,
  ``Complexity Growth with Lifshitz Scaling and Hyperscaling Violation,''
  JHEP {\bf 1807}, 042 (2018)
  doi:10.1007/JHEP07(2018)042
  [arXiv:1802.06740 [hep-th]].
  %%CITATION = doi:10.1007/JHEP07(2018)042;%%
  %35 citations counted in INSPIRE as of 19 Mar 2019


%\cite{Gouteraux:2011ce}\cite{Huijse:2011ef}{Alishahiha:2012qu}
\bibitem{Gouteraux:2011ce}
  B.~Gouteraux and E.~Kiritsis,
  ``Generalized Holographic Quantum Criticality at Finite Density,''
  JHEP {\bf 1112} (2011) 036
  [arXiv:1107.2116 [hep-th]].
  %%CITATION = ARXIV:1107.2116;%%


%\cite{Huijse:2011ef}
\bibitem{Huijse:2011ef} 
  L.~Huijse, S.~Sachdev and B.~Swingle,
  ``Hidden Fermi surfaces in compressible states of gauge-gravity duality,''
  arXiv:1112.0573 [cond-mat.str-el].
  %%CITATION = ARXIV:1112.0573;%%


%\cite{Alishahiha:2012qu}
\bibitem{Alishahiha:2012qu} 
  M.~Alishahiha, E.~O Colgain and H.~Yavartanoo,
  ``Charged Black Branes with Hyperscaling Violating Factor,''
  JHEP {\bf 1211}, 137 (2012)
  [arXiv:1209.3946 [hep-th]].
  %%CITATION = ARXIV:1209.3946;%%
  %20 citations counted in INSPIRE as of 16 Dec 2013



%\cite{Skenderis:2002wp}
\bibitem{Skenderis:2002wp} 
  K.~Skenderis,
  ``Lecture notes on holographic renormalization,''
  Class.\ Quant.\ Grav.\  {\bf 19}, 5849 (2002)
  doi:10.1088/0264-9381/19/22/306
  [hep-th/0209067].
  %%CITATION = doi:10.1088/0264-9381/19/22/306;%%
  %803 citations counted in INSPIRE as of 12 Aug 2018

%\cite{Kiritsis:2015doa}
\bibitem{Kiritsis:2015doa} 
  E.~Kiritsis and Y.~Matsuo,
  ``Charge-hyperscaling violating Lifshitz hydrodynamics from black-holes,''
  JHEP {\bf 1512}, 076 (2015)
  doi:10.1007/JHEP12(2015)076
  [arXiv:1508.02494 [hep-th]].
  %%CITATION = doi:10.1007/JHEP12(2015)076;%%
  %22 citations counted in INSPIRE as of 24 Jan 2019


%\cite{Cardy:2018jho}
\bibitem{Cardy} 
  J.~Cardy,
  ``$T\overline T$ deformations of non-Lorentz invariant field theories,''
  arXiv:1809.07849 [hep-th].
  %%CITATION = ARXIV:1809.07849;%%
  %12 citations counted in INSPIRE as of 26 May 2019

\bibitem{Lloyd:2000}
S.~ Lloyd, ``Ultimate physical limits to computation,''  Nature {\bf 406} (2000) 1047,
[arXiv:quant-ph/9908043]


%\cite{Brown:2015bva}\cite{Brown:2015lvg}
\bibitem{Brown:2015bva}
  A.~R.~Brown, D.~A.~Roberts, L.~Susskind, B.~Swingle and Y.~Zhao,
  ``Holographic Complexity Equals Bulk Action?,''
  Phys.\ Rev.\ Lett.\  {\bf 116}, no. 19, 191301 (2016)
  doi:10.1103/PhysRevLett.116.191301
  [arXiv:1509.07876 [hep-th]].
  %%CITATION = doi:10.1103/PhysRevLett.116.191301;%%
  %43 citations counted in INSPIRE as of 17 Dec 2016

 %\cite{Brown:2015lvg}
\bibitem{Brown:2015lvg}
  A.~R.~Brown, D.~A.~Roberts, L.~Susskind, B.~Swingle and Y.~Zhao,
  ``Complexity, action, and black holes,''
  Phys.\ Rev.\ D {\bf 93}, no. 8, 086006 (2016)
  doi:10.1103/PhysRevD.93.086006
  [arXiv:1512.04993 [hep-th]].
  %%CITATION = doi:10.1103/PhysRevD.93.086006;%%
  %28 citations counted in INSPIRE as of 17 Dec 2016





%\cite{Parattu:2015gga}
\bibitem{Parattu:2015gga} 
  K.~Parattu, S.~Chakraborty, B.~R.~Majhi and T.~Padmanabhan,
  ``A Boundary Term for the Gravitational Action with Null Boundaries,''
  Gen.\ Rel.\ Grav.\  {\bf 48}, no. 7, 94 (2016)
  doi:10.1007/s10714-016-2093-7
  [arXiv:1501.01053 [gr-qc]].
  %%CITATION = doi:10.1007/s10714-016-2093-7;%%
  %74 citations counted in INSPIRE as of 04 Nov 2018

%\cite{Parattu:2016trq}
\bibitem{Parattu:2016trq} 
  K.~Parattu, S.~Chakraborty and T.~Padmanabhan,
  ``Variational Principle for Gravity with Null and Non-null boundaries: A Unified Boundary Counter-term,''
  Eur.\ Phys.\ J.\ C {\bf 76}, no. 3, 129 (2016)
  doi:10.1140/epjc/s10052-016-3979-y
  [arXiv:1602.07546 [gr-qc]].
  %%CITATION = doi:10.1140/epjc/s10052-016-3979-y;%%
  %25 citations counted in INSPIRE as of 04 Nov 2018



%\cite{Lehner:2016vdi}
\bibitem{Lehner:2016vdi}
  L.~Lehner, R.~C.~Myers, E.~Poisson and R.~D.~Sorkin,
  ``Gravitational action with null boundaries,''
  Phys.\ Rev.\ D {\bf 94} (2016) no.8,  084046
  doi:10.1103/PhysRevD.94.084046
  [arXiv:1609.00207 [hep-th]].
  %%CITATION = doi:10.1103/PhysRevD.94.084046;%%
  %16 citations counted in INSPIRE as of 17 Dec 2016



%\cite{Carmi:2017jqz}
\bibitem{Carmi:2017jqz} 
  D.~Carmi, S.~Chapman, H.~Marrochio, R.~C.~Myers and S.~Sugishita,
  ``On the Time Dependence of Holographic Complexity,''
  JHEP {\bf 1711}, 188 (2017)
  doi:10.1007/JHEP11(2017)188
  [arXiv:1709.10184 [hep-th]].
  %%CITATION = doi:10.1007/JHEP11(2017)188;%%
  %18 citations counted in INSPIRE as of 02 Feb 2018





%\cite{Reynolds:2016rvl}
\bibitem{Reynolds:2016rvl} 
  A.~Reynolds and S.~F.~Ross,
  ``Divergences in Holographic Complexity,''
  Class.\ Quant.\ Grav.\  {\bf 34}, no. 10, 105004 (2017)
  doi:10.1088/1361-6382/aa6925
  [arXiv:1612.05439 [hep-th]].
  %%CITATION = doi:10.1088/1361-6382/aa6925;%%
  %49 citations counted in INSPIRE as of 31 Dec 2018
  
 
%\cite{Stanford:2014jda}
\bibitem{Stanford:2014jda} 
  D.~Stanford and L.~Susskind,
  ``Complexity and Shock Wave Geometries,''
  Phys.\ Rev.\ D {\bf 90}, no. 12, 126007 (2014)
  doi:10.1103/PhysRevD.90.126007
  [arXiv:1406.2678 [hep-th]].
  %%CITATION = doi:10.1103/PhysRevD.90.126007;%%
  %254 citations counted in INSPIRE as of 21 Aug 2019



\end{thebibliography}
\end{document}